\documentclass[a4paper,11pt]{article}
\pdfoutput=1 
\usepackage{authblk}
\usepackage[T1]{fontenc}
\usepackage{graphicx}
\usepackage{caption,subcaption}
\usepackage[export]{adjustbox}
\usepackage{babel}
\usepackage{amsmath,amssymb,xcolor}
\pdfoutput=1 
\usepackage{jheppub}
\usepackage{blindtext}
\usepackage{hyperref}
\usepackage{url}
\usepackage{xspace}
\allowdisplaybreaks

\usepackage[normalem]{ulem}
\usepackage{multirow}
\usepackage{bashful}
\usepackage{eufrak}
\usepackage{todonotes}
\usepackage{caption}
\usepackage{subcaption}
\usepackage{comment}
\usepackage{graphbox}

\newcommand{\e}{\ensuremath{\epsilon}}

\newcommand{\inlinegraphics}[2][]{\ensuremath{\vcenter{\hbox{\includegraphics[#1]{#2}}}}}

\title
{Two-loop differential cross section for $l^{+}l^{-}\rightarrow\gamma\gamma$ with power corrections in the lepton mass}
\author{Vyacheslav A. Stotsky}
\affiliation{Budker Institute of Nuclear Physics, 630090, Novosibirsk, Russia}
\emailAdd{stotsky.slava@gmail.com}

\abstract{
	We calculate the two-loop QED amplitude for $l^{+}l^{-}\to\gamma\gamma$ and the corresponding unpolarized differential cross section in the high-energy region, retaining power corrections in the lepton mass. For both this process and the inverse process $\gamma\gamma\to l^{+}l^{-}$, these corrections can be important for accurate predictions when the lepton mass cannot be neglected, particularly in the near-threshold region. The amplitude is presented in a tensor basis and can therefore be evaluated for arbitrary polarizations of the external particles. All results are expressed in terms of classical polylogarithms $\mathrm{Li}_n$ for efficient numerical evaluation. Numerical tests indicate that the truncated expansions remain accurate for $s,|t|,|u|\gtrsim40m_l^2$. At leading power in the lepton mass, the two-loop amplitude agrees with the prediction of QED mass factorization.
}

\begin{document}
	\maketitle
	\flushbottom

	\section{Introduction}
		Future electron–positron colliders and existing high-precision experiments require increasingly precise theoretical calculations. In particular, experiments at FCC-ee are expected to require theoretical results with an accuracy of better than $0.1\%$ \cite{freitas2019theoretical}. To achieve such precision for QED processes, next-to-next-to-leading-order (NNLO) calculations are generally needed. Basic QED processes are used for luminosity determination, radiative-correction studies, and background simulation. Their theoretical uncertainties therefore directly affect the systematic uncertainties of experimental measurements. For example, \texttt{MCGPJ} \cite{fedotovich2010monte} and \texttt{BabaYaga@NLO} \cite{balossini2008photon} quote theoretical accuracies of $0.2\%$ and $0.1\%$, respectively. Both generators combine NLO calculations with higher-order contributions implemented using approximate resummation methods. 
		
		These quoted accuracies usually refer to integrated cross sections or sufficiently inclusive observables and do not necessarily apply to differential distributions. For example, in the CMD-3 measurement of the $e^{+}e^{-} \rightarrow \pi^{+}\pi^{-}$ cross section \cite{ignatov2024measurement}, the modified \texttt{MCGPJ} and \texttt{BabaYaga@NLO} generators gave predictions differing by about $10\%$ in the far tails of the lepton momentum spectra. This difference illustrates the possible systematic uncertainty associated with the predictions for differential cross sections. More accurate NNLO QED results and their implementation in event generators can help reduce systematic uncertainties in current and future collider experiments.
		
		The inverse processes, including $\gamma\gamma\to e^{+}e^{-}$ and $\gamma\gamma\to\mu^{+}\mu^{-}$, are also of considerable interest. Such processes are studied, for example, to test QED in photon--photon interactions. Electron--positron production has been measured by STAR \cite{adam2021measurement}, while muon-pair production has been studied at HERA and by CMS and ATLAS \cite{aktas2004muon,cms2012exclusive,aad2015measurement}. The HERA, CMS, and ATLAS measurements cover photon--photon center-of-mass energies well above the corresponding pair-production thresholds, whereas the STAR measurement probes comparatively low pair invariant masses. Other experimental programs explore energies closer to threshold. Design studies for MeV photon--photon colliders consider electron--positron production near threshold \cite{takahashi2018light,zhou2026estimation}, while possible RHIC measurements of muon-pair production have been discussed for pair invariant masses approximately two to three times the threshold value $2m_\mu$ \cite{shao2023azimuthal}. In this regime, the lepton mass cannot be neglected. Away from the immediate threshold region, power corrections in $m_l^2/s$ can be sizable and should be included in theoretical predictions.
		
		NNLO calculations of QED $2\to2$ processes with massive electrons are considerably more complicated than the corresponding massless calculations because the finite electron mass must be retained as an additional physical scale. In massless QED, the scattering amplitude contains collinear divergences associated with the emission of a photon collinear with an electron or positron. For a nonzero electron mass, these collinear divergences appear as logarithms of the energy-to-mass ratio. An exact treatment of the electron mass can lead to non-polylogarithmic functions and makes the calculations of $2\rightarrow2$ processes significantly more difficult.
		
		One approach is therefore to treat the electron mass as small compared with the characteristic energy scales. In that case, differential equations yielding non-polylogarithmic contributions can be solved near their singular points using the Frobenius method. The solutions of such equations have the form of a generalized power series in the electron mass. Their coefficients can usually be expressed in terms of Goncharov's polylogarithms, as was done, for example, in Refs.~\cite{gerasimov2025electron,delto2023two}. We use this approach in the present work.
		
		An alternative approach is to use QED mass factorization. The massive amplitude is then related to the corresponding massless amplitude up to power-suppressed terms in the electron mass. This approach has been used in calculations of the differential cross sections for  $e^{+} e^{-} \rightarrow e^{+} e^{-}$ \cite{becher2007two}, $e^{+} e^{-} \rightarrow \gamma \gamma^{*}$ \cite{fadin2023two} and $e^{+}e^{-} \rightarrow 2\gamma$ \cite{engel2026low,naterop2021electron}. The general validity and region of applicability of this relation, however, require further investigation, and direct comparisons with explicit massive calculations are desirable. In the present work, we perform such a comparison for the $l^{+}l^{-} \rightarrow2\gamma$ amplitude at two loops. To the best of our knowledge, this is the first direct test of the QED mass-factorization formula for a $2\rightarrow2$ process at the two-loop level. Further details are given in Section~\ref{sec:factorization}.
		
		The purpose of this work is to calculate the two-loop QED amplitudes for $l^{+}l^{-}\to\gamma\gamma$ with arbitrary external polarizations and the corresponding unpolarized differential cross section. We work in the high-energy, large-angle region and retain power corrections in the lepton mass. For definiteness, we present the results for $e^{+}e^{-}\rightarrow\gamma\gamma$ and denote the electron mass by $m_e$. The corresponding results for other lepton flavors are obtained by replacing $m_e$ with the appropriate lepton mass.
		
		We introduce the conventional Mandelstam variables:
		\begin{align}
			s=(p_1+p_2)^2,  \quad  t = &(p_1 - k_1)^2, \quad u=(p_1 - k_2)^2 = 2 m_e^2-t-s, \nonumber\\ 
			p_1^2 = & \ p_2^2 = m_e^2, \quad k_1^2 = k_2^2 = 0,  
		\end{align}
		where $p_1$ and $p_2$ are the electron and positron momenta, respectively, $k_1$ and $k_2$ are the photon momenta, and $m_e$ is the electron mass. We consider the kinematic region 
		\begin{equation}
			s, \, |t|, \, |u| \gg m_e^2.
		\end{equation}	
		The physical region is defined by
		\begin{equation}
			s > 4m_e^2, \qquad
			-\frac{s}{4}\left(1+\sqrt{1-\frac{4m_e^2}{s}}\right)^2
			\le t \le
			-\frac{s}{4}\left(1-\sqrt{1-\frac{4m_e^2}{s}}\right)^2 .
			\label{reg}
		\end{equation}
		For convenience, we introduce dimensionless variables
		\begin{align}
			m= \frac{m_e}{\sqrt{s}}, \quad T = m^2 - \frac{t}{s}, \quad U = m^2 - \frac{u}{s} = 1 - T, \quad \beta = \sqrt{1-4m^2}, \quad \cos\theta = \frac{1-2T}{\beta}.
		\end{align}
		In terms of these variables, the physical region \eqref{reg} becomes
		\begin{equation}
			m^2 < \frac{1}{4}, \qquad  \frac{1-\beta}{2} \leq T \leq \frac{1+ \beta}{2}. 
		\end{equation}
		The paper is organized as follows. Section~\ref{sec:ampcalc} describes the calculation of the amplitude and differential cross section. Section~\ref{sec:cs} presents the results and examines the region of applicability of the expansion. In Section~\ref{sec:factorization}, we compare our results with known calculations and with the prediction of QED mass factorization.
		
	\section{Details of the calculation}
	\label{sec:ampcalc}
		\subsection{Diagrams and amplitude}
			\subsubsection{Master integrals}
			The main input for the amplitude calculation is the set of two-loop massive master integrals evaluated in Ref.~\cite{lee2024master}. That reference provides their mass expansions through order $m^{12}$. We use the same expansion accuracy throughout the present calculation. We summarize the relevant results below. The calculation uses eight integral families, denoted by $\texttt{gg}_1,\dots,\texttt{gg}_8$ and shown in Fig.~\ref{fig:diagrams}. Systems of differential equations with respect to $m^2$ and $\tau=-t/s$ were constructed for $165$ master integrals:
			\begin{align}
				\partial_{m^2} \boldsymbol{j} &=M_{m} (m^2,\tau,\epsilon) \, \boldsymbol{j}, \label{dem1}\\
				\partial_{\tau} \boldsymbol{j} &= M_{\tau} (m^2,\tau,\epsilon) \, \boldsymbol{j},  \label{det1}
			\end{align}
			where $\boldsymbol{j}$ is the vector of master integrals and $M_{m}$ and $M_{\tau}$ are rational matrices. For substitution of the master integrals into the amplitude, it is very convenient to change variables from $\tau$ to
			\begin{equation}
				T = m^2 + \tau.
			\end{equation}
			Starting from the results of Ref.~\cite{lee2024master}, we re-expand the master integrals in the electron mass at fixed $T$. The resulting generalized power series have the form
			\begin{equation}
				\boldsymbol{j}(m^2,T,\epsilon) = \biggr[ \sum_{i=1}^{7}  (m^2)^{\lambda_i} \sum_{n=0}^{6} \sum_{k=0}^{K_i} B_{i,n,k} \cdot m^{2n} \cdot \frac{\ln^{k} m^2}{k!} \biggr] \cdot \widetilde{\boldsymbol{C}} (T,\epsilon), \label{gps1}
			\end{equation}
			where $\widetilde{\boldsymbol{C}}$ is a vector whose components are $\epsilon$-series of uniform transcendental weight containing Goncharov’s  polylogarithms with argument $T$ and indices $0$ and $1$, $B_{i,n,k}$ are matrix coefficients exact in $\epsilon$. The exponents and logarithmic degrees are
			\begin{align}
				\{\lambda_{1},\,\lambda_{2},\,\dots,\,\lambda_{7} \}  & = \{0,-\frac{1}{2}  -  3\epsilon,-\frac{1}{2}-2\epsilon,-4\epsilon,-3\epsilon,-2\epsilon,-\epsilon\}, \label{eigenv}\\
				\{K_{1},K_{2},\dots,K_{7} \} & =  \{0,0,0,0,0,2,1\}.
			\end{align}
			The resulting expressions for the master integrals at fixed $T$ are available in the GitHub repository \url{https://github.com/VStotsky/eegg_highenergy/}.
			
			Since the unrenormalized two-loop amplitude contains poles up to order $\epsilon^{-4}$, it is sufficient to retain terms through transcendental weight $4$ in the uniform-weight vector $\widetilde{\boldsymbol{C}}$ of Ref.~\cite{lee2024master}. For the terms through weight $4$, we rewrite these Goncharov's polylogarithms in terms of classical polylogarithms $\operatorname{Li}_n$.
			
			\subsubsection{Diagram construction and renormalization}
\begin{figure}
					\begin{subfigure}{\textwidth}
						\includegraphics[width=\textwidth]{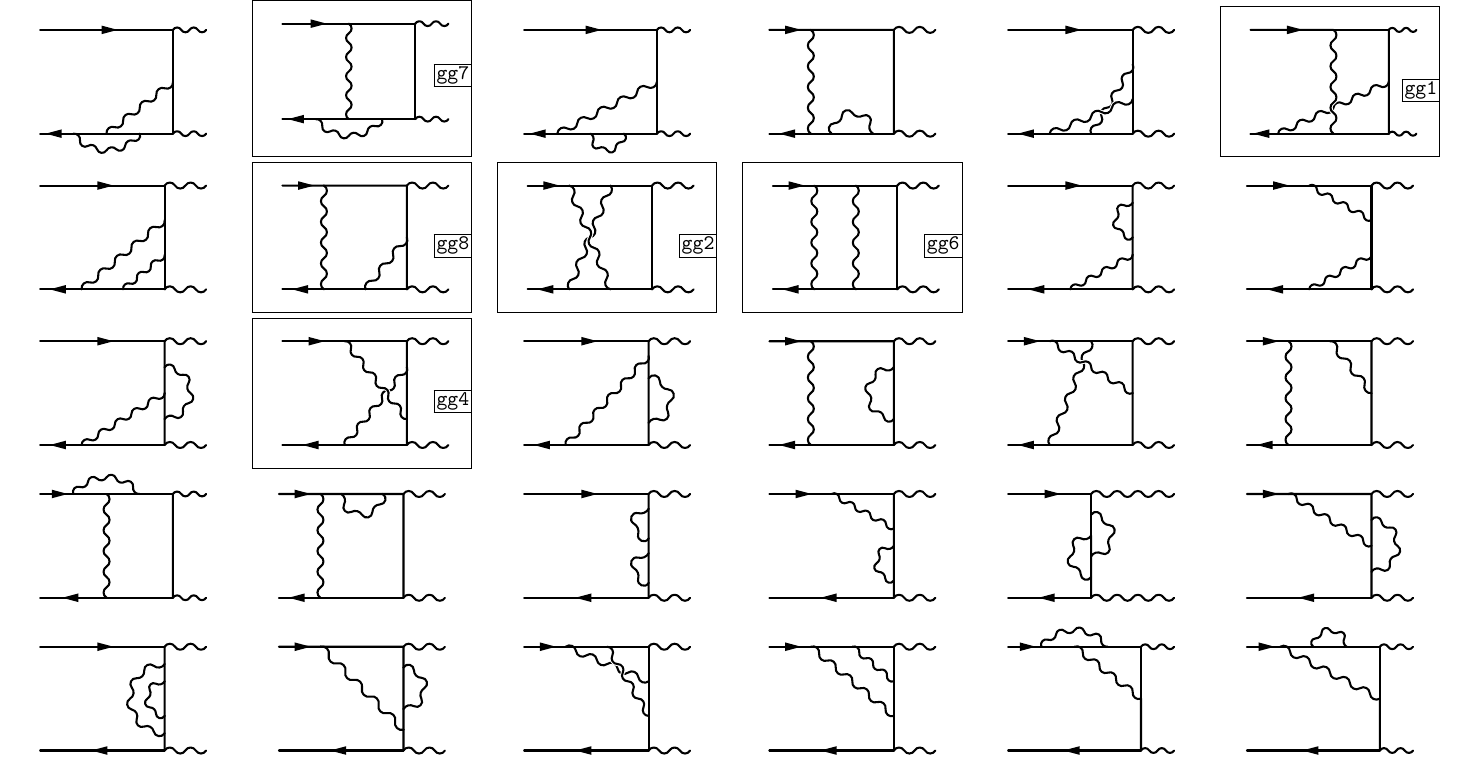}
						\caption{Pure photonic diagrams.}
						\label{fig:ph}
					\end{subfigure}
					\begin{subfigure}{\textwidth}
						\centering\includegraphics[width=0.66\textwidth]{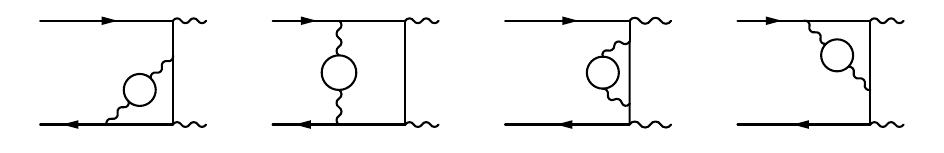}
						\caption{Diagrams with a vacuum-polarization insertion.}
						\label{fig:po}
					\end{subfigure}
					\begin{subfigure}{\textwidth}
						\centering\includegraphics[width=0.44\textwidth]{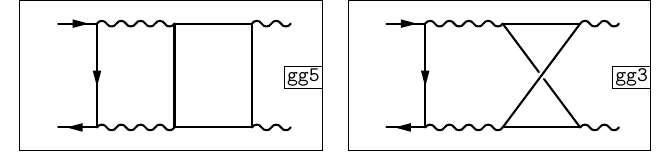}
						\caption{Diagrams with a light-by-light insertion.}
						\label{fig:lbl}
					\end{subfigure}
					\caption{Two-loop diagrams for the process $e^+ e^- \rightarrow \gamma\gamma$.}
					\label{fig:diagrams}
				\end{figure}

			We first construct the unrenormalized amplitude. Its two-loop part consists of $74$ diagrams, half of which are shown in Fig.~\ref{fig:diagrams}. The remaining diagrams are obtained by exchanging the two final-state photons, which corresponds to $t\leftrightarrow u$, or $T\leftrightarrow U=1-T$.  When this symmetry is expressed in terms of $T$ and $U$, no additional expansion of the polylogarithms in the electron mass is required.
			
			We generate an expression in momentum space for each diagram using the \texttt{QGRAF} package \cite{qgraf}. Each expression is then written in the form of a linear combination of integrals from the families $\texttt{gg}_1,\dots,\texttt{gg}_8$. IBP-reduction was done using the \texttt{Sparx} package \cite{LeeSparx}. At the time of writing, \texttt{Sparx} is unpublished; its public release as an add-on to \texttt{LiteRed2}~\cite{lee2013litered} is planned.

			To construct the $e^{+}e^{-} \rightarrow \gamma\gamma$ amplitude, we use the on-shell renormalization scheme. Let us write the amplitude as a series in $a_0=\alpha_0/(4\pi)$, where $\alpha_0$ is the bare fine-structure constant:
			\begin{equation}
				A = A_{0} + a_{0} A_{1} + a_{0}^{2} A_{2} + \mathcal{O}(a_{0}^3). \label{unramp}
			\end{equation}
			The right-hand side of this equation is also expressed in terms of the bare mass $m_{0}$ and the bare electron field $\psi_{0}$. To obtain the renormalized amplitude, we express the bare quantities in terms of on-shell ones as follows
			\begin{align}
				&m_0 = m_e Z_{m}^{\mathrm{OS}}, \nonumber \\*
				&\psi_{0} = \psi \sqrt{Z_{2}^{\mathrm{OS}}} , \nonumber \\*
				& a_0 = a \big(Z_{3}^{\mathrm{OS}} \big)^{-1} , \label{renorm0}
			\end{align}
			where 
			\begin{equation}
				a_0 = \frac{\alpha_{0}}{4\pi}, \quad a = \frac{\alpha}{4\pi}. \label{renorm1}
			\end{equation}
			We expand the renormalization constants in $a_0$ as
			\begin{equation}
				 Z_{\mathrm{x}}^{\mathrm{OS}} = 1 + \sum_{l\geq1} \biggr(a_0 \frac{\Gamma(1+\e)}{(4\pi)^{-\e}m_e^{2\e}} \biggr)^{l} Z_{\mathrm{x},l}^{\mathrm{OS}} , \label{renorm2}
			\end{equation}
			where $Z_{\mathrm{x}}^{\mathrm{OS}}$ denotes any of $Z_m^{\mathrm{OS}}$, $Z_2^{\mathrm{OS}}$ and $Z_3^{\mathrm{OS}}$. Substituting Eqs.~\eqref{renorm0} and \eqref{renorm2} into the amplitude generates the counterterms. 
			
\begin{figure}
					\includegraphics[width=\textwidth]{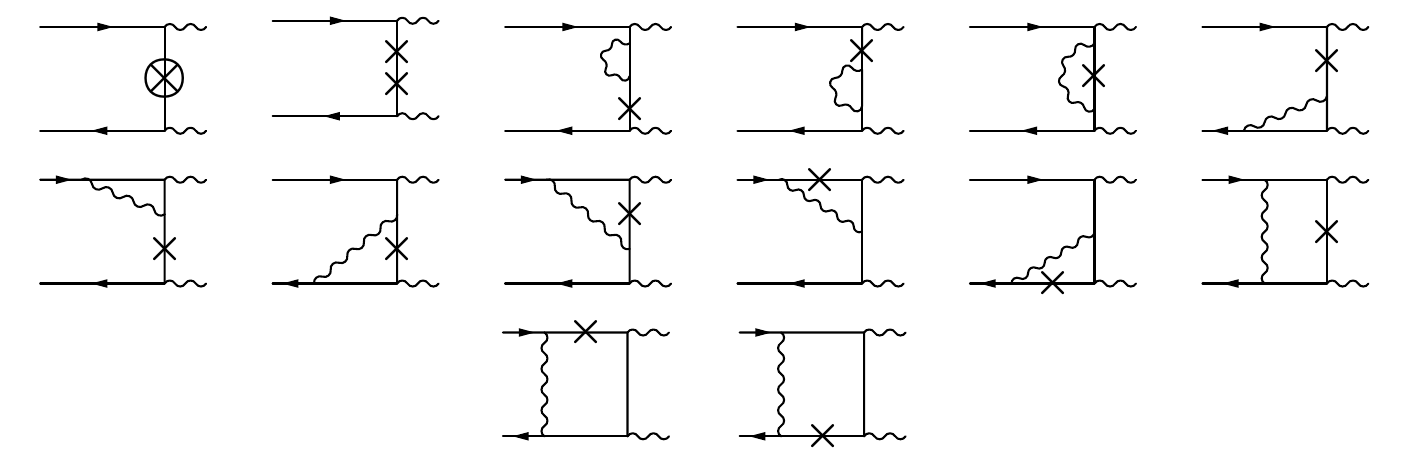}
					\caption{Two-loop counterterms containing $Z_{m,2}^{\mathrm{OS}}$ and $Z_{m,1}^{\mathrm{OS}}$.}
					\label{fig:ct}
			\end{figure}

			The two‑loop counterterms containing $Z_m^{\mathrm{OS}}$ are shown in Fig.~\ref{fig:ct}. The cross and circled cross denote the following vertices:
			\begin{align}
				&\inlinegraphics[height=1.5cm]{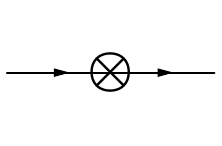} = m_e a^{2} \bigg(Z_{m,2}^{\mathrm{OS}} - Z_{3,1}^{\mathrm{OS}} Z_{m,1}^{\mathrm{OS}} \bigg) \biggr(\frac{\Gamma(1+\e)}{(4\pi)^{-\e}m_{e}^{2\e}}\biggr)^2 , \label{zm2}\\*
				& \inlinegraphics[height=1.5cm]{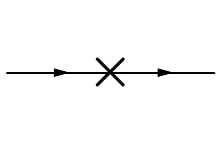} =  m_e a Z_{m,1}^{\mathrm{OS}} \bigg( \frac{\Gamma(1+\e)}{(4\pi)^{-\e}m_{e}^{2\e}} \bigg). \label{zm1}  
			\end{align}
			Here
			\begin{equation}
				Z_{m,1}^{\mathrm{OS}} = Z_{2,1}^{\mathrm{OS}}= - \frac{3-2\e}{\e(1-2\e)}, \qquad Z_{3,1}^{\mathrm{OS}} = - \frac{4 n_f}{3 \epsilon}.
			\end{equation}
			Expressions for the renormalization constants $Z_{m}^{\mathrm{OS}}$, $Z_{2}^{\mathrm{OS}}$ and $Z_{3}^{\mathrm{OS}}$ with two-loop accuracy can be found in Refs.~\cite{broadhurst1991gauge,melnikov2000three,fadin2023two}. The explicit $\epsilon$-series for $Z_{m,2}^{\mathrm{OS}}$ and $Z_{2,2}^{\mathrm{OS}}$ used in this work are presented in Appendix~\ref{app:renorm}.
			
			The two-loop counterterms, which are proportional to $Z_{3,2}^{\mathrm{OS}}$, $ Z_{2,2}^{\mathrm{OS}}$, $ (Z_{2,1}^{\mathrm{OS}})^{2}$ and $Z_{2,1}^{\mathrm{OS}} \times Z_{m,1}^{\mathrm{OS}}$ are obtained by multiplying the Born or one-loop amplitude by the corresponding factors and therefore do not require a separate calculation.
			
			\subsubsection{Soft factorization}
			
			We denote the renormalized amplitude by $A_r$ and expand it as
			\begin{equation}
				A_r = A_{r,0} + a A_{r,1} + a^2 A_{r,2} + \mathcal{O}(a^3). \label{Ar}
			\end{equation}
			Soft photon emission leads to infrared divergences, which we factorize following Ref.~\cite{yennie1961infrared}. The unpolarized differential cross section, averaged over the initial spins and summed over the final photon polarizations, takes the form
			\begin{equation}
				d \sigma = \frac{1}{64 \pi} |A_r|^2 \frac{-s \cdot dT}{I^2} = \frac{1}{64 \pi} e^{a \mathcal{W}(\omega_0) + 2a \Re\mathcal{V}} \left|\mathcal{H}\right|^2 \frac{-s \cdot dT}{I^2} , \label{ds0}
			\end{equation}  
			where $I^2=\frac{s^2}{4}(1-4m^2)$ and $\mathcal{H}$ is the hard amplitude,
			\begin{equation}
			 	\mathcal{H} = \mathcal{H}_0 + a \mathcal{H}_1 + a^2 \mathcal{H}_2 + \mathcal{O}(a^3),
			\end{equation}
			which is related to $A_{r}$ by
			\begin{equation}
				\mathcal{H} = e^{-a \mathcal{V}} A_{r}. \label{HviaA}
			\end{equation}
			The definitions of $\mathcal{V}$ and $\mathcal{W}$, together with explicit expressions through $\mathcal{O}(\epsilon)$, can be found in Ref.~\cite{gerasimov2025electron}. For $e^{+}e^{-} \rightarrow \gamma\gamma$, their small-mass expansions read
			\begin{align}
				&\mathcal{V} = -\frac{2(1+L_m + i\pi) }{\e} + L_m^{2} +L_m -2 + \frac{4\pi^2}{3} - i\pi  \nonumber \\*
				& \qquad - \biggr(\frac{4}{\e} + 2(L_m + 3i\pi -1) \biggr)m^2 +\dots+ \mathcal{O}(m^{14}),  \label{Vseries}\\ 
				& \mathcal{W}(\omega_0) = \frac{4(1 + L_m)}{\e} - 2 \biggr(L_{m}^{2} + 2 L_m + 4 L_{m}L_{\omega_0} + 4 L_{\omega_0} + \frac{2\pi^2}{3} \biggr) + \nonumber\\
				& \qquad + \biggr(\frac{8}{\e} -8(3 L_m + 2 L_{\omega_0}) \biggr)m^2 +\dots+ \mathcal{O}(m^{14}),  \label{Wseries}
			\end{align}
			where $L_{m} = \ln (m^2)$ and $L_{\omega_0} = \ln (\frac{2\omega_0}{\sqrt{s}})$, and $\omega_0$ is the soft-photon energy cutoff. Substituting Eqs.~\eqref{Vseries} and \eqref{Ar} into \eqref{HviaA} and expanding the exponential through order $a^2$, we find that the $\epsilon$-pole terms cancel at each order in electron mass considered. The \e-poles also cancel in the exponential factor of the differential cross section \eqref{ds0}:
			\begin{align}
				\mathcal{W}(\omega_0) + 2 \Re\mathcal{V} &= -2(2+ L_m + 4 L_{\omega_0} + 4 L_m L_{\omega_0} - \frac{2\pi^2}{3}) - \nonumber\\ & - 4 m^2 (-1+7L_m + 4 L_{\omega_0}) + \dots + \mathcal{O}(m^{14}), \label{w2v}
			\end{align}
			Therefore, we can set $\epsilon=0$ in the expression for $\mathcal{H}$.
			
		\subsection{Invariant amplitudes}
			To calculate the cross section for arbitrary polarizations of the external particles, we first determine the independent tensor structures that can appear in the amplitude. The coefficients of these structures, called invariant amplitudes, depend only on the kinematic variables and the electron mass. Once these amplitudes are known, they can be used to construct the cross section and to study its expansion in the electron mass. Accordingly, we decompose the hard amplitude  \eqref{HviaA} as follows:
			\begin{equation}
				\mathcal{H} = \sum_{k} \mathcal{H}_{k} \bar{v} (p_2) \, \Gamma^{\mu \nu}_{k} \, u (p_1) e^{*}_{\mu} (k_2) e^{*}_{\nu} (k_1), \label{amp1}
			\end{equation}
			where $\Gamma^{\mu \nu}_{k}$ are tensor structures built from external momenta and $\gamma$ matrices, and $\mathcal{H}_k$ are the corresponding invariant amplitudes, which depend on the variables $m$, $T$ and $U$. Here, $u$ and $v$ are the electron and positron bispinors, and $e(k_{1,2})$ are the photon polarization vectors. We choose a complete, linearly independent set of basis tensors that are separately transverse, so that gauge invariance is manifest:
			\begin{equation}
				\Gamma^{\mu \nu}_{k}  k_{2 \mu} = \Gamma^{\mu \nu}_{k}  k_{1 \nu} = 0, \label{id1}
			\end{equation}
			C-parity and P-parity conservation yields the following constraints, respectively:
			\begin{align}
				& \Gamma^{\mu \nu}_{k} (k_1,k_2,T,U) =  \gamma^2 \gamma^0 \left[\Gamma_k^{\mu\nu}(k_1,k_2,U,T)\right]^{\mathrm T} \gamma^2 \gamma^0, \nonumber\\*
				& \Gamma^{\mu \nu}_{k} (k_1,k_2) = P^{\mu}_{\mu'}P^{\nu}_{\nu'} \gamma^{0} \Gamma^{\mu' \nu'}_{k} (k_1^{P},k_2^{P}) \gamma^{0}, \label{cp}
			\end{align}
			where $P^\mu{}_\nu=\operatorname{diag}(1,-1,\dots,-1)$, $(k_{1,2}^P)^\mu=P^\mu{}_\nu k_{1,2}^\nu
			=(k_{1,2}^0,-\vec k_{1,2})$. For convenience, we additionally choose structures symmetric under interchange of the two photons:
			\begin{equation}
				\Gamma^{\mu \nu}_{k} (k_1,k_2,T,U) = \Gamma^{\nu \mu}_{k} (k_2,k_1,U,T), \label{bose}
			\end{equation}
			The corresponding invariant amplitudes are then symmetric under $T \leftrightarrow U$.
	
			In $d$ dimensions we choose the following eight independent structures:
			\begin{align}
				&	\Gamma^{\mu \nu}_{1} = g^{\mu \nu}_{\perp} , \quad \Gamma^{\mu \nu}_{2} = k_{2 \perp}^{\nu} k_{1 \perp}^{\mu}, \quad \Gamma^{\mu \nu}_{3} = (T-\frac{1}{2})(\gamma_{\perp}^{\nu} k_{1 \perp}^{\mu} - \gamma_{\perp}^{\mu} k_{2 \perp}^{\nu}), \nonumber \\ 
				& \Gamma^{\mu \nu}_{4} = (T-\frac{1}{2}) \hat{K} g^{\mu \nu}_{\perp}, \quad \Gamma^{\mu \nu}_{5} =  (T-\frac{1}{2})\hat{K} k_{2 \perp}^{\nu} k_{1 \perp}^{\mu} , \quad \Gamma^{\mu \nu}_{6} = (T-\frac{1}{2})(\gamma_{\perp}^{\nu} \gamma_{\perp}^{\mu} - \gamma_{\perp}^{\mu} \gamma_{\perp}^{\nu}) \nonumber \\ 
				& \Gamma^{\mu \nu}_{7} =  (T-\frac{1}{2}) \big[ (\hat{K} \gamma_{\perp}^{\nu} - \gamma_{\perp}^{\nu}\hat{K}) k_{1 \perp}^{\mu} + (\hat{K} \gamma_{\perp}^{\mu} - \gamma_{\perp}^{\mu}\hat{K}) k_{2 \perp}^{\nu} + \frac{8(m^2-T+T^2)}{(2T-1)^3} \Gamma^{\mu \nu}_{6} \big], \nonumber\\
				& \Gamma^{\mu\nu}_{8} = \gamma_{\perp}^{\mu} \hat{K} \gamma_{\perp}^{\nu} - \gamma_{\perp}^{\nu} \hat{K} \gamma_{\perp}^{\mu} + \frac{4 m}{(2T-1)^2}  \Gamma^{\mu \nu}_{6}, \label{struct}
			\end{align}
			where 
			\begin{align}
				g^{\mu \nu}_{\perp} = g^{\mu \nu} - \frac{k_{1}^{\mu}k_{2}^{\nu}}{(k_1 \cdot k_2)}, &\quad k_{1 \perp}^{\mu} = k_{1}^{\mu} - \frac{(k_1 \cdot k_2) Q^{\mu}}{(k_{2} \cdot Q)}, \quad k_{2 \perp}^{\nu} = k_{2}^{\nu} - \frac{(k_1 \cdot k_2) Q^{\nu}}{(k_{1} \cdot Q)}, \nonumber\\ \gamma_{\perp}^{\mu} =& \gamma^{\mu} - \frac{\hat{k}_{2} k_{1}^{\mu}}{(k_1 \cdot k_2)}, \qquad \gamma_{\perp}^{\nu} = \gamma^{\nu} - \frac{\hat{k}_{1} k_{2}^{\nu}}{(k_1 \cdot k_2)}, \nonumber\\
				&\hat{K} = \gamma_\alpha (k_1^\alpha - k_2^\alpha), \quad Q = p_1 - p_2.
			\end{align}
			
			It is instructive to compare this $d$-dimensional basis with the independent helicity amplitudes in four dimensions. In $d=4$, the last two structures in Eq.~\eqref{struct} do not contribute: their contractions with the external spinors and polarization vectors vanish by virtue of four-dimensional Dirac-algebra identities. The number of remaining structures coincides with the number of helicity amplitudes $M_{\lambda_{e^{+}}\lambda_{e^{-}}\lambda_{\gamma_1}\lambda_{\gamma_2}}$, taking into account the above-described symmetries:
			\begin{align}
				M_{++++}, \qquad M_{+++-}, \qquad M_{++--}, \nonumber\\
				M_{+-++}, \qquad M_{+-+-}, \qquad M_{+---}. 
			\end{align}
			In the massless limit, helicity conservation forbids the amplitudes in the first row. At finite electron mass, they are therefore power-suppressed, leaving three independent contributions at leading power. In our choice of structures \eqref{struct}, the expansion of the invariant amplitudes $\mathcal{H}_1$, $\mathcal{H}_2$, $\mathcal{H}_6$, and $\mathcal{H}_7$ start at $\mathcal{O}(m)$, so these amplitudes vanish as  $m \rightarrow 0$. At leading power in the electron mass, the hard amplitude therefore reduces to
			\begin{equation}
				\mathcal{H} =  \bar{v} (p_2) \, (\mathcal{H}_{3}\Gamma^{\mu \nu}_{3}+\mathcal{H}_{4}\Gamma^{\mu \nu}_{4}+\mathcal{H}_{5}\Gamma^{\mu \nu}_{5}) \, u (p_1) e^{*}_{\mu} (k_2) e^{*}_{\nu} (k_1).
			\end{equation}
			
			To extract the invariant amplitudes, we define the projected tensor structures 
			\begin{equation}
				\mathcal{G}_{l} = \overline{v} (p_2) \, \Gamma^{\alpha \beta}_{l} \, u (p_1) e^{*}_{\alpha} (k_2) e^{*}_{\beta} (k_1).
			\end{equation} 
			For these structures, we introduce the inner product, denoted by $\langle \dots \rangle$:
			\begin{equation}
				\langle\mathcal{G}_k,\mathcal{G}_l^*\rangle
				=\operatorname{Tr}\left[
				\frac{\hat p_2-m_e}{2}\Gamma_k^{\mu\nu}
				\frac{\hat p_1+m_e}{2}\Gamma_{l,\mu\nu}^*\right].
				\label{matr}
			\end{equation}
			The brackets include averaging over the initial-particle spins and summing over the final-photon polarizations. We define the elements of the $8\times8$ Gram matrix by $G_{lk}=\langle\mathcal{G}_k,\mathcal{G}_l^*\rangle$. Projecting Eq.~\eqref{amp1} onto $\mathcal{G}_l$ then gives
			\begin{equation}
				\langle\mathcal{H},\mathcal{G}_l^*\rangle
				=\sum_{k=1}^{8}\mathcal{H}_kG_{lk}. \label{amp2}
			\end{equation}
			The invariant amplitudes are obtained by inverting the Gram matrix:
			\begin{equation}
				\mathcal{H}_k=\sum_{l=1}^{8}(G^{-1})_{kl}
				\langle\mathcal{H},\mathcal{G}_l^*\rangle. \label{Hk}
			\end{equation}

			For practical purposes, we rescale the invariant amplitudes according to
			\begin{equation}
				\mathrm{H}_l=\mathcal{D}_{lk}\mathcal{H}_k, \label{newH}
			\end{equation} 
			where
			\begin{equation}
				\mathcal{D} = (m^2 - (1-T)T) \cdot \operatorname{diag}\bigg(\frac{1}{m},\frac{1}{m},1,1,1,\frac{1}{m},\frac{1}{m},1\bigg). \label{support}
			\end{equation}
			With this rescaling, the power expansions of the invariant amplitudes $\mathrm{H}_k$ in $m$ start at $\mathcal{O}(m^0)$. The overall factor $m^2-(1-T)T$ is chosen because successive terms in the mass expansion contain progressively higher powers of $T(1-T)$ in their denominators, a feature whose origin is discussed in Section~\ref{sec:cs}.
			
			To complete the picture, we now express the unpolarized differential cross section in terms of the invariant amplitudes. We write each $\mathrm{H}_k$ as an expansion in $a=\frac{\alpha}{4\pi}$:
			\begin{equation}
				\mathrm{H}_k = \mathrm{H}_{0,k} + a \mathrm{H}_{1,k} + a^2 \mathrm{H}_{2,k} + \mathcal{O} (a^3). \label{hviaa}
			\end{equation}
			The Born amplitudes $\mathrm{H}_{0,k}$ are real in our conventions. Substituting Eqs.~\eqref{amp1} and \eqref{newH} into Eq.~\eqref{ds0}, we obtain
			\begin{align}
				d\sigma  = \frac{-s \, dT}{64 \pi I^2} e^{a \mathcal{W}(\omega_0) + 2a \Re\mathcal{V}} (\mathcal{D}^{-1})_{lq} G_{qp} (\mathcal{D}^{-1})_{pk} \bigg[&\mathrm{H}_{0,k}\mathrm{H}_{0,l} + 2a \mathrm{H}_{0,k} \Re\mathrm{H}_{1,l} + \nonumber\\ & + a^2 (\mathrm{H}_{1,k}\mathrm{H}_{1,l}^{*} + 2\mathrm{H}_{0,k} \Re\mathrm{H}_{2,l} )  + \mathcal{O} (a^3) \bigg], \label{ds00}
			\end{align}
			where summation over repeated indices is implied. Equation~\eqref{ds00} retains the soft factor in exponentiated form. To obtain the fixed-order cross section through order $a^2$, this exponential must also be expanded through the same order.
			
			The invariant amplitudes \eqref{hviaa}, the combination $\mathcal{W}+2\Re\mathcal{V}$ \eqref{w2v}, and the unpolarized differential cross section \eqref{cs1} are provided in the ancillary files described in Section~\ref{sec:ancillary}.
	\section{Cross section and region of applicability}	
	\label{sec:cs}
			Substituting the invariant amplitudes obtained above into Eq.~\eqref{ds00} and expanding the result in the electron mass, we obtain the differential cross section in the form
			\begin{equation}
				d \sigma = d \sigma_0 (1 + \delta_1 + \delta_2 + \mathcal{O}(a^3)), \label{cs1}
			\end{equation}
			where $d \sigma_0$ is the Born cross section, and $\delta_1$ and $\delta_2$ are the one- and two-loop corrections, respectively. Their dependence on the logarithms  $L_m = \ln m^2 = \ln(m_e^2/s)$ and $L_{\omega_0} = \ln \frac{2\omega_0}{\sqrt{s}}$ has the following form:
			\begin{align}
				& \frac{\delta_1}{a} = -8 L_{m}L_{\omega_0} - 8 L_{\omega_0} - 6 L_m + A_{0} + m^2 \biggr(\sum_{k=0}^{2}\sum_{l=0}^{1}A_{2,kl}L_{m}^{k} L_{\omega_0}^{l} \biggr) + \dots + \mathcal{O}(m^{14}), \label{d1form} \\
				&\frac{\delta_2}{a^2} = \frac{8}{9} L_m^{3}  + \sum_{k,l=0}^{2} B_{0,kl} L_{m}^{k} L_{\omega_0}^{l} + m^2 \biggr( \sum_{k=0}^{4} \sum_{l=0}^{2} B_{2,kl} L_m^{k}L_{\omega_0}^{l} \biggr) + \dots +\mathcal{O}(m^{13}). \label{d2form}
			\end{align}
			The coefficients $A_0$ and $A_{i,kl}$ are rational functions of $T$ multiplied by the logarithms of $T$ and $1-T$, whereas the coefficients $B_{i,kl}$ are rational functions of $T$ multiplied by classical polylogarithms $\operatorname{Li}_s$ with $s\le4$ and arguments $T$, $1-T$ and $\frac{T}{T-1}$.
			
			The one- and two-loop corrections have different mass expansions. The one-loop correction contains only even powers of $m$, whereas the two-loop correction also includes odd powers, starting at $m^3$. The origin of these odd-power contributions is discussed in the next section.
			
\begin{figure}
				\centering
				\begin{subfigure}{0.45\textwidth}
					\centering
					\includegraphics[scale=0.33]{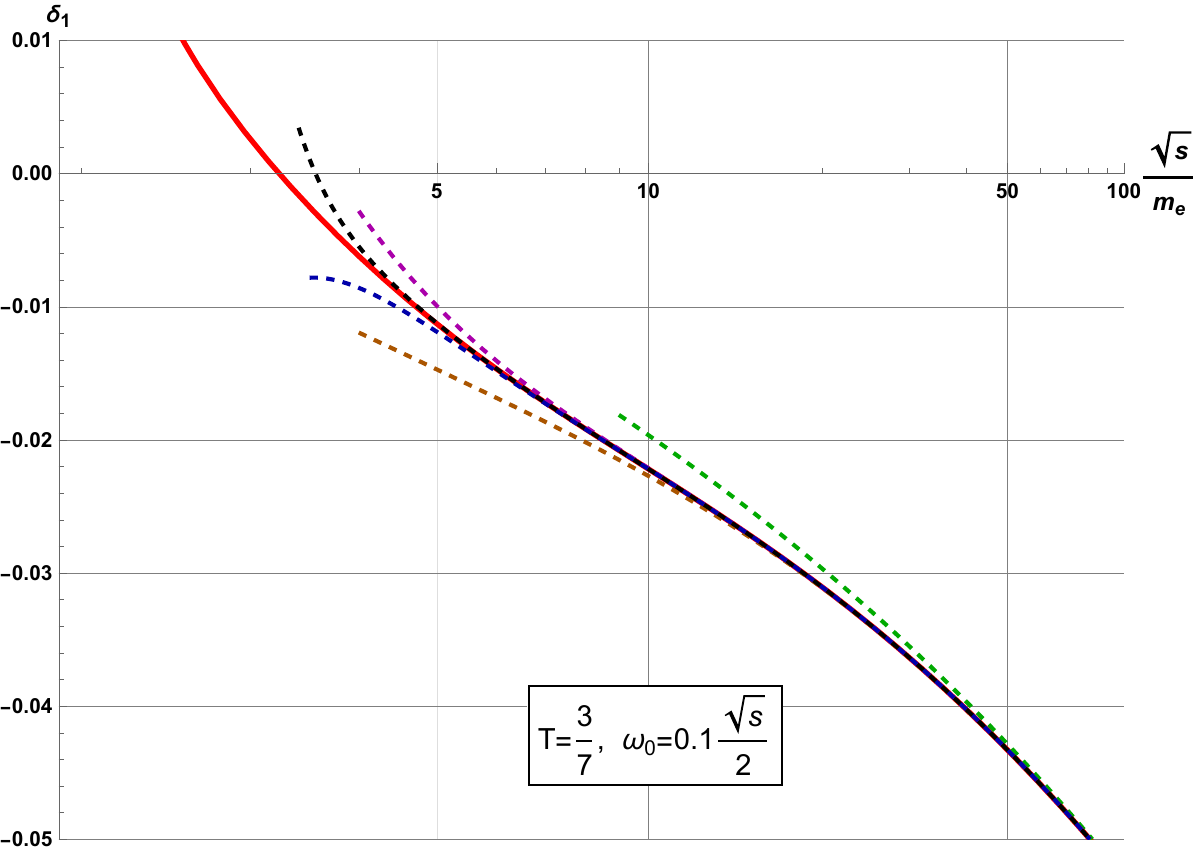}
					\caption{Dependence of $\delta_1$ on energy at $T=\frac{3}{7}$.}
					\label{fig:delta1s}
				\end{subfigure}
				\begin{subfigure}{0.45\textwidth}
					\centering
					\includegraphics[scale=0.33]{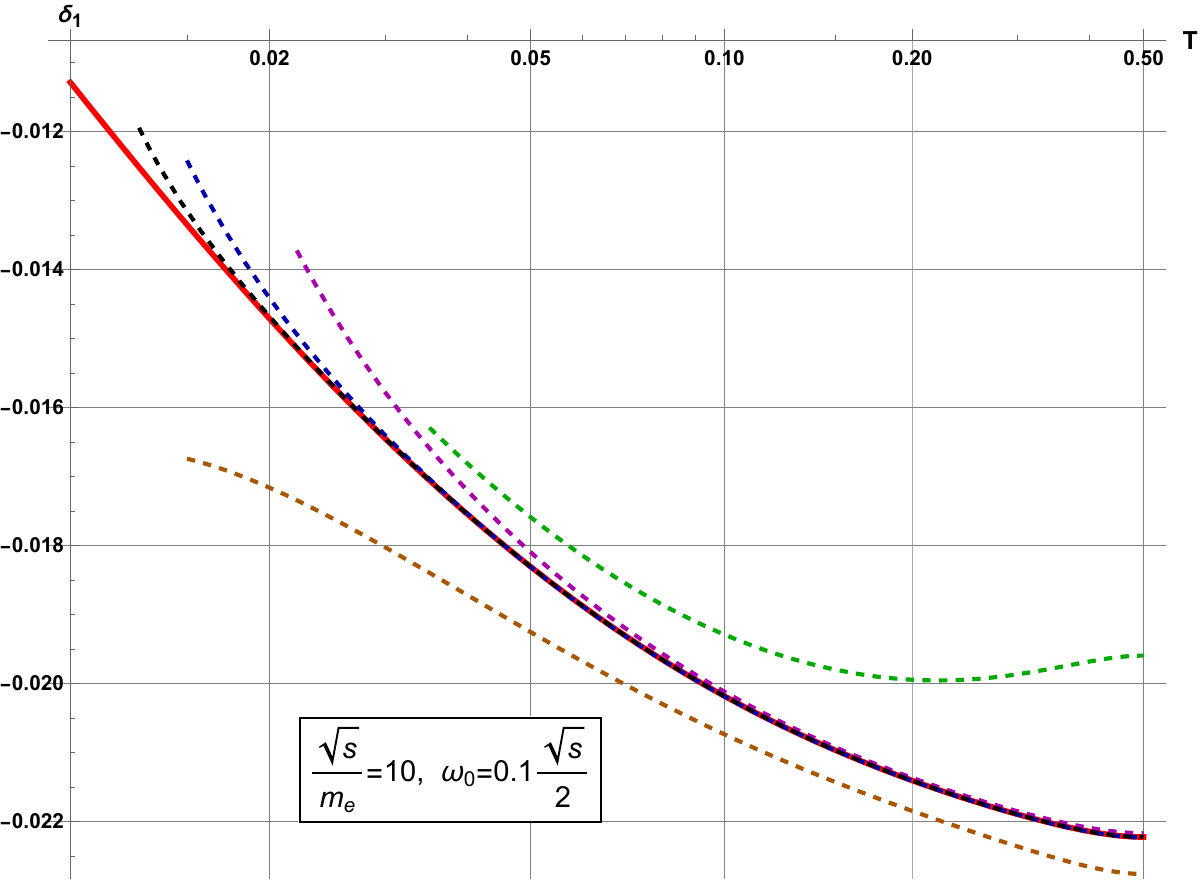}
					\caption{Dependence of $\delta_1$ on $T$ at $\sqrt{s}=10 m_e$.}
					\label{fig:delta1T}
				\end{subfigure}
				\caption{One-loop correction $\delta_1$ at $\omega_0=0.1\sqrt{s}/2$. The red solid line shows the result exact in $m$ \cite{berends1973hard}. The green, orange, magenta, blue, and black dashed lines show successive approximations with remainders $\mathcal{O}(m^2)$, $\mathcal{O}(m^4)$, $\mathcal{O}(m^6)$, $\mathcal{O}(m^8)$, and $\mathcal{O}(m^{14})$, respectively.}
				\label{delta1}
			\end{figure}

			With these expansions in hand, we examine their numerical applicability. We begin with the one-loop correction, for which an exact massive result is available in Ref.~\cite{berends1973hard}. Figures~\ref{fig:delta1s} and \ref{fig:delta1T} show the dependence of $\delta_1$ on $\sqrt{s}/m_e$ and $T$, respectively. In both figures, the red curve shows the result exact in the electron mass, the green dashed curve shows the massless limit, and the other curves show approximations with successively more power corrections. For a generic value of $T$ not too close to $0$ or $1$, the truncated one-loop result remains accurate down to approximately $\sqrt{s}\gtrsim4m_e$, as shown in Fig.~\ref{fig:delta1s}. For small $T$, corresponding to small scattering angles, higher energies are required to maintain the same accuracy. Figure~\ref{fig:delta1T} further shows that, even at the relatively low energy $\sqrt{s}=10m_e$, the highest available truncation is indistinguishable from the exact massive result at the scale of the figure for $T\gtrsim10^{-2}$.
			
\begin{figure}
				\begin{subfigure}{\textwidth}
					\centering
					\includegraphics[scale=0.5]{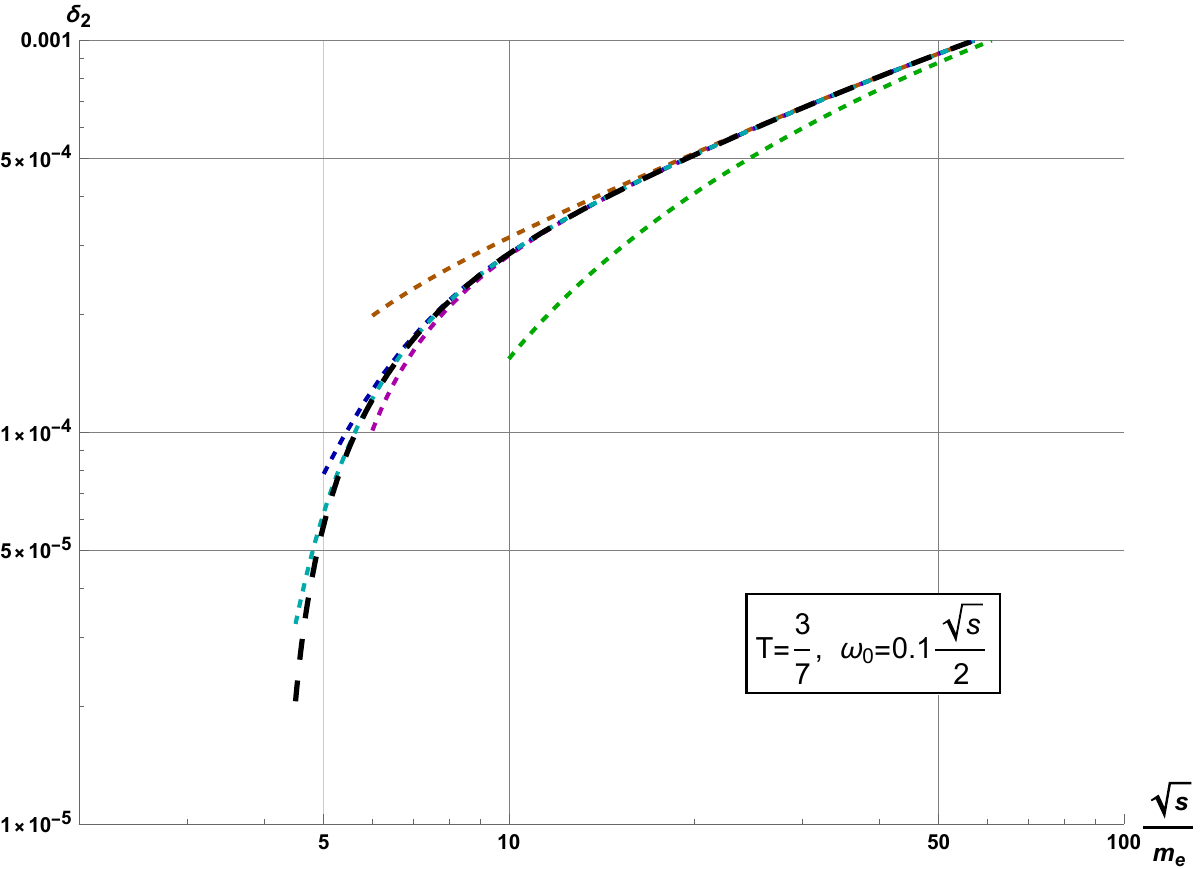}
					\caption{Dependence of $\delta_2$ on energy at $T=\frac{3}{7}$.}
					\label{fig:delta2le}
				\end{subfigure}
				\centering
				\begin{subfigure}{0.45\textwidth}
					\centering\includegraphics[scale=0.33]{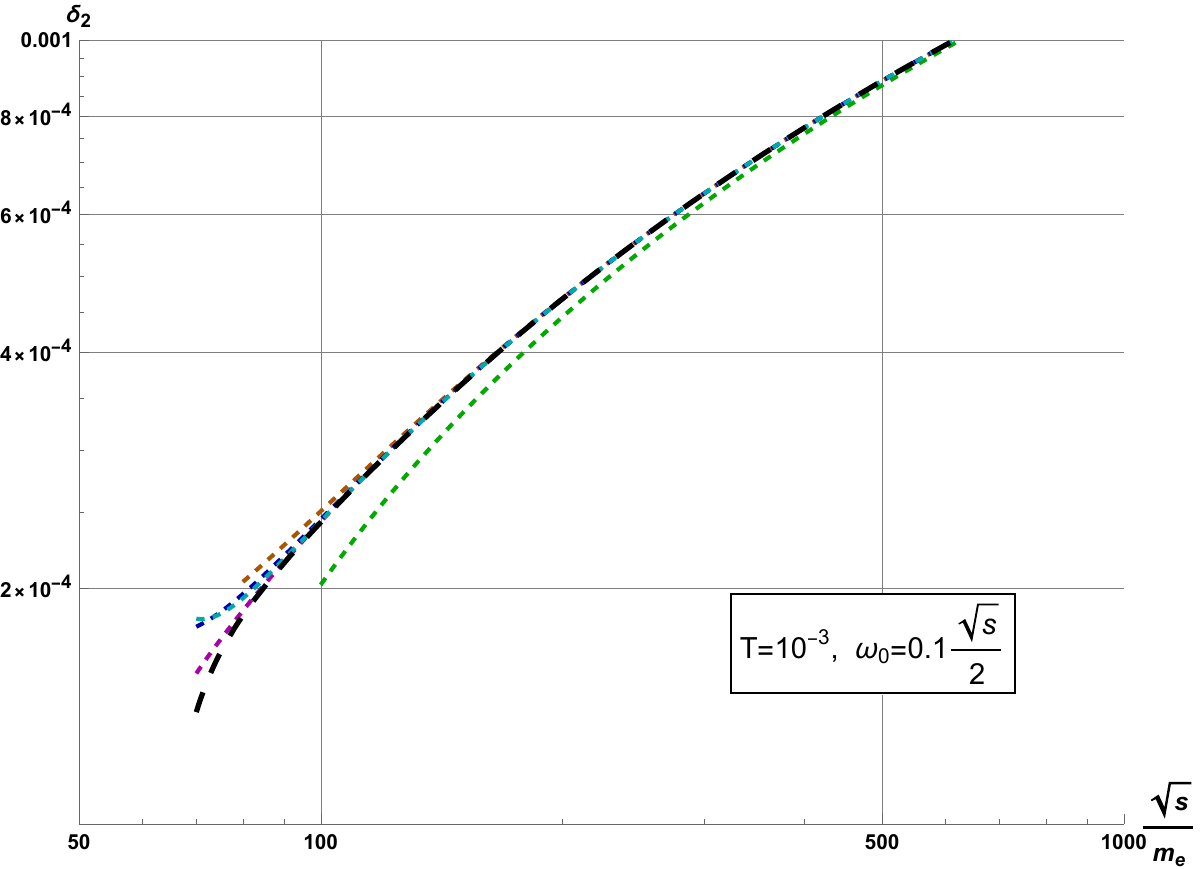}
					\caption{Dependence of $\delta_2$ on energy at $T=10^{-3}$.}
					\label{fig:delta2spt}
				\end{subfigure}
				\begin{subfigure}{0.45\textwidth}
					\centering\includegraphics[scale=0.33]{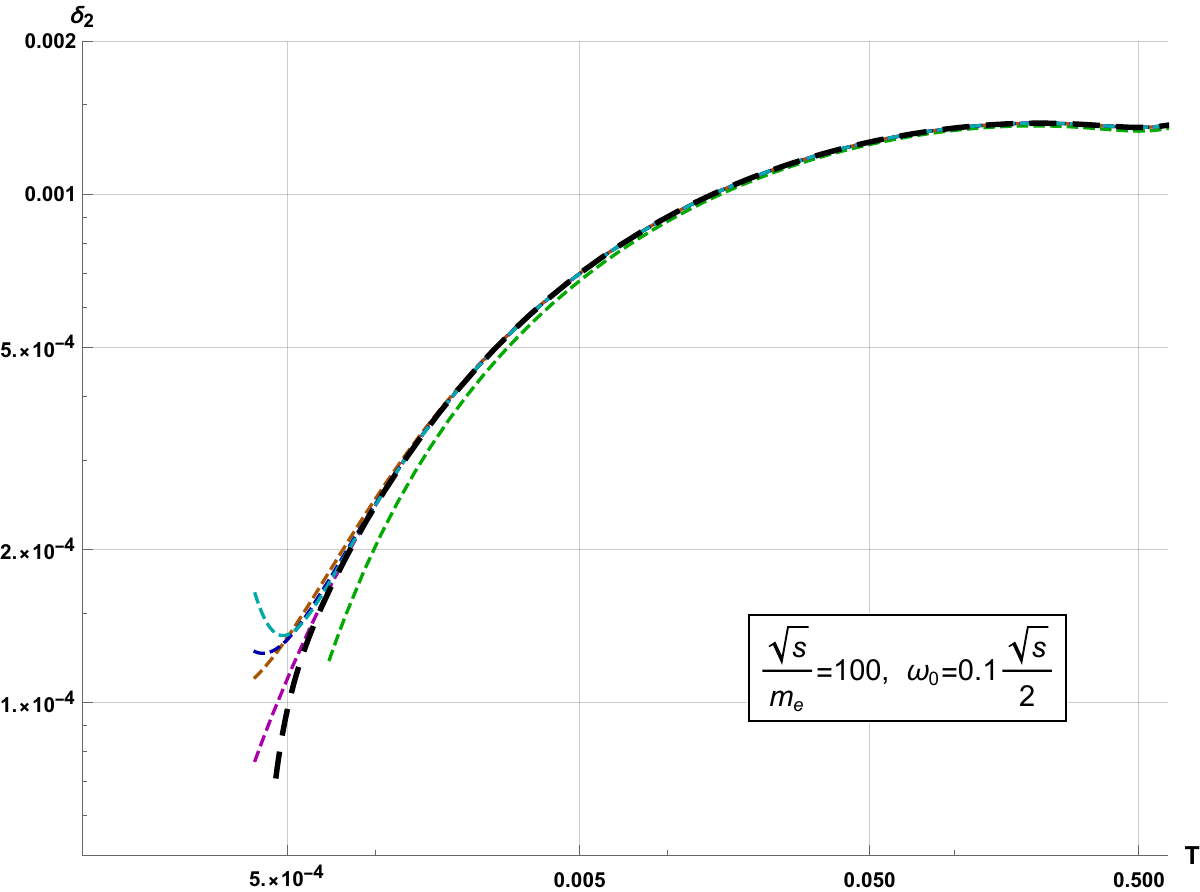}
					\caption{Dependence of $\delta_2$ on $T$ at $\sqrt{s}/m_e = 100$.}
					\label{fig:delta2Tpt}
				\end{subfigure}
				\caption{Two-loop correction $\delta_2$ at $\omega_0=0.1 \frac{\sqrt{s}}{2}$. The green, orange, magenta, blue, cyan, and black dashed lines show successive approximations with remainders $\mathcal{O}(m^1)$, $\mathcal{O}(m^3)$, $\mathcal{O}(m^5)$, $\mathcal{O}(m^7)$, $\mathcal{O}(m^{9})$, and $\mathcal{O}(m^{13})$, respectively.}
				\label{delta2pt}
			\end{figure}

			A similar analysis can be carried out for the two-loop correction, with the results shown in Figs.~\ref{fig:delta2le}, \ref{fig:delta2spt}, and \ref{fig:delta2Tpt}. Since no two-loop result exact in the electron mass is available, we assess the convergence of the expansion by comparing successive truncations and use the highest-order available truncation as a reference. For a generic value of $T$, Fig.~\ref{fig:delta2le} shows good agreement between successive truncations down to approximately $\sqrt{s}\gtrsim6m_e$. For the small value $T=10^{-3}$, Fig.~\ref{fig:delta2spt} shows that successive approximations remain close to the reference down to $\sqrt{s}\approx100m_e$. Similarly, Fig.~\ref{fig:delta2Tpt} indicates that the expansion remains close to the reference down to $T\approx10^{-3}$ at $\sqrt{s}/m_e=100$.
			
			The one- and two-loop results show the same loss of accuracy near $T=0$ and $T=1$. This behavior follows from the fact that the mass expansion is performed at fixed $T$ and is not uniform near these endpoints. The coefficients of successive terms contain inverse powers of $T$ and $1-T$, so that the effective expansion parameters in these regions are $m^2/T$ and $m^2/(1-T)$, respectively. Since $|t|=sT-m_e^2$ and $|u|=s(1-T)-m_e^2$, the expansion requires $|t|\gg m_e^2$ and $|u|\gg m_e^2$ in addition to $s\gg m_e^2$. Therefore, the parametric region of applicability is
			\begin{equation}
				T\gg\frac{m_e^2}{s}, \qquad
				1-T\gg\frac{m_e^2}{s}.
			\end{equation}
			Numerically, however, the deepest available truncations can remain accurate close to the boundaries suggested by these parametric estimates:
			\begin{equation}
				T\gtrsim\frac{m_e^2}{s}, \qquad
				1-T\gtrsim\frac{m_e^2}{s}. \label{limit1}
			\end{equation}
			In terms of the scattering angle, these conditions become
			\begin{equation}
				\theta\gtrsim\frac1\gamma, \qquad
				\pi-\theta\gtrsim\frac1\gamma, \label{const1}
			\end{equation}
			where
			\begin{equation}
				\cos\theta=\frac{1-2T}{\beta}, \qquad
				\beta=\sqrt{1-4m^2}, \qquad
				\gamma=\frac{1}{2m}.
			\end{equation}
			
			The energy scan at fixed generic $T$ tests the expansion away from the endpoints, while the scans at small $T$ and near $T=1$ probe the additional conditions $|t|\gg m_e^2$ and $|u|\gg m_e^2$. Combining these numerical tests, we conservatively estimate that, for the truncation orders considered here, the expansion is applicable for
			\begin{equation}
				s,|t|,|u|\gtrsim40m_e^2.
			\end{equation}
			
			In the kinematic range examined here, the leading-power approximation alone is not sufficiently accurate for either the one- or two-loop correction, whereas including the first power correction substantially improves the accuracy.
			
\begin{figure}
				\centering\includegraphics[width=0.8\textwidth]{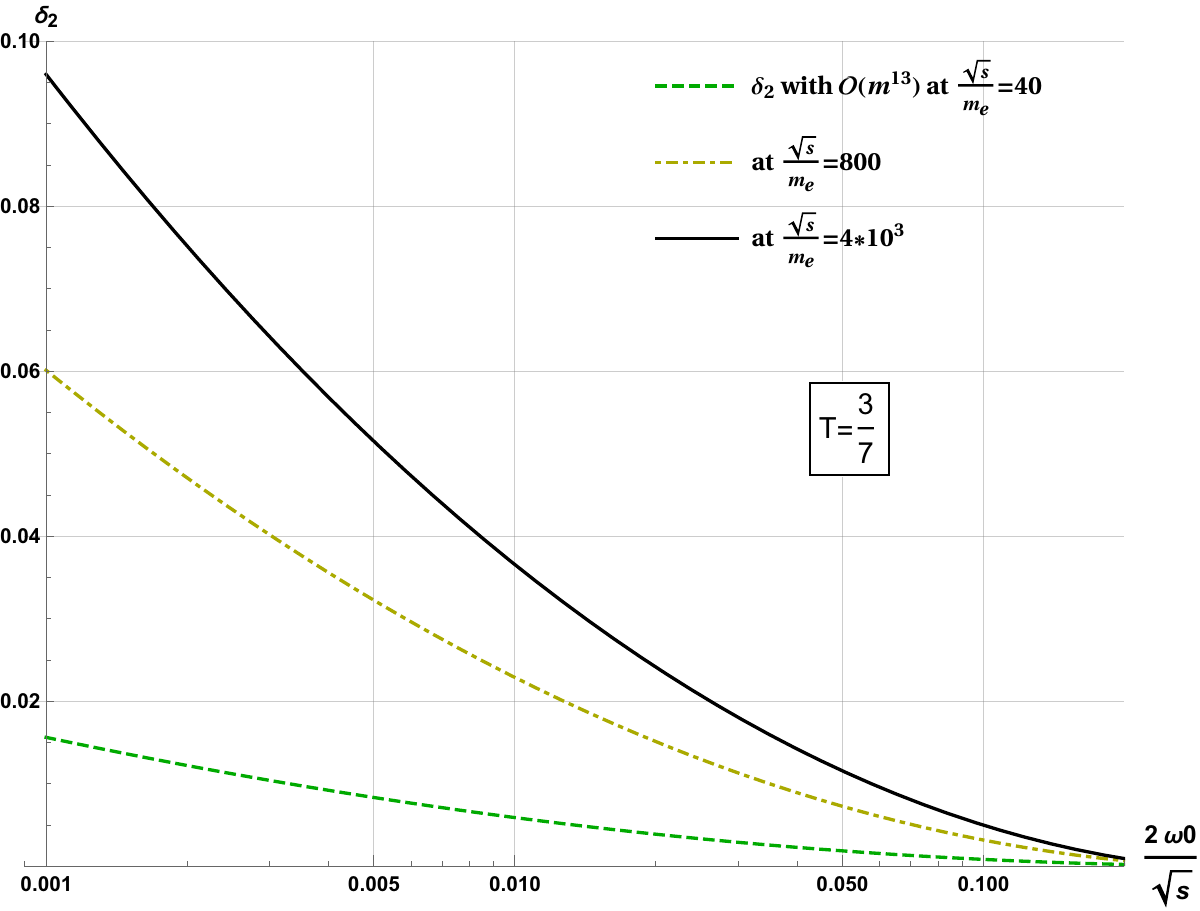}
				\caption{Dependence of $\delta_2$ on $\frac{2\omega_0}{\sqrt{s}}$ at $T=\frac{3}{7}$.}
				\label{fig:omega0}
			\end{figure}

			The numerical estimates given above were obtained for $2\omega_0/\sqrt{s}=1/10$. It is also instructive to consider lower values of this ratio. For the first few terms of the mass expansion, the agreement between the truncated results and the reference result improves substantially as $2\omega_0/\sqrt{s}$ decreases. The highest-order truncation, by contrast, is much less sensitive to $\omega_0$, and the corresponding numerical applicability estimates remain essentially unchanged. The corrections themselves nevertheless depend on $\omega_0$, with $\delta_1$ increasing and $\delta_2$ decreasing as $\omega_0$ increases. The dependence of $\delta_2$ on $\omega_0$ is shown in Fig.~\ref{fig:omega0}.
			
\begin{figure}
				\centering\includegraphics[width=0.8\textwidth]{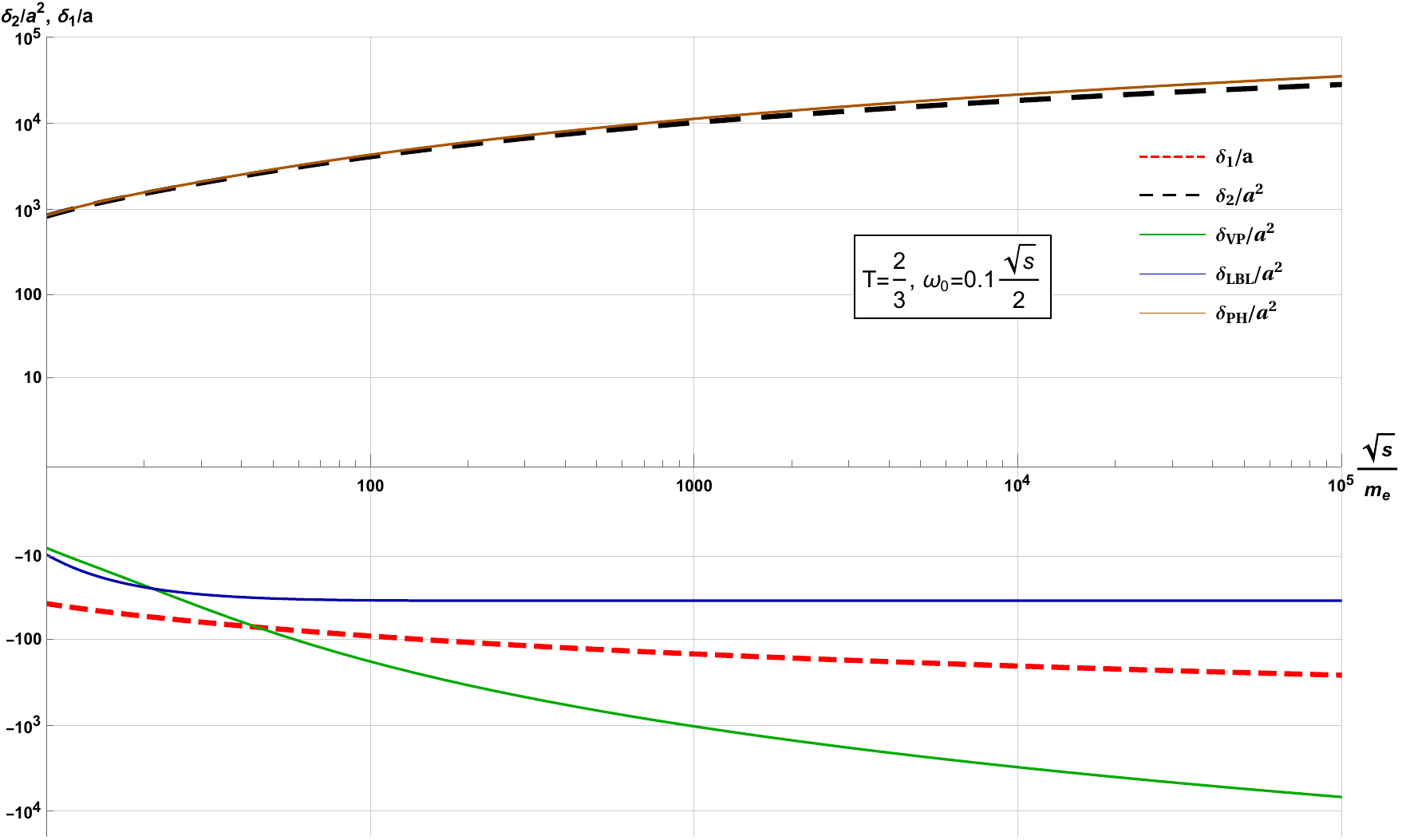}
				\caption{Dependence of $\delta_2$, $\delta_1$ and individual contributions $\delta_{\mathrm{PH}}$ (Fig.~\ref{fig:ph}), $\delta_{\mathrm{VP}}$ (Fig.~\ref{fig:po}) and $\delta_{\mathrm{LBL}}$ (Fig.~\ref{fig:lbl}) on $\sqrt{s}/m_e$ at $T=\frac{2}{3}$, $\omega_0 = 0.1 \frac{\sqrt{s}}{2}$.}
				\label{fig:delta12s}
			\end{figure}

			Finally, we consider the high-energy behavior of the two-loop correction and its individual diagrammatic contributions. Figure~\ref{fig:delta12s} displays the highest-order available truncation of $\delta_2/a^2$, with remainder $\mathcal{O}(m^{13})$, as a black dashed line and the one-loop correction $\delta_1/a$ as a red solid line. It also shows the individual two-loop contributions from the pure photonic diagrams in Fig.~\ref{fig:ph}, the vacuum-polarization diagrams in Fig.~\ref{fig:po}, and the light-by-light diagrams in Fig.~\ref{fig:lbl} in orange, green, and blue lines, respectively.
			
			The high-energy behavior can be understood from the leading logarithms. The high-energy growth of the pure photonic contribution is dominated by the term quadratic in $L_m$:
			\begin{equation}
				\delta_{\text{PH}}/a^2 \overset{\frac{\sqrt{s}}{m_e}\rightarrow \infty}{=} \frac{1}{2} L_m^{2} (6+8 L_{\omega_0})^2 + \mathcal{O}(L_m).
			\end{equation}
			The light-by-light contribution $\delta_{\mathrm{LBL}}$ remains finite in the massless limit and is numerically small throughout the energy range shown. The vacuum-polarization contribution $\delta_{\mathrm{VP}}$, on the other hand, becomes dominant only at extremely high energies, far beyond the experimentally relevant range considered here:
			\begin{equation}
				\delta_{\mathrm{VP}} /a^2 \overset{\frac{\sqrt{s}}{m_e}\rightarrow \infty}{=} \frac{8 L_m^{3}}{9} + \frac{76 L_m^{2}}{9} + \mathcal{O}(L_m).
			\end{equation}
			Moreover, throughout the energy range considered, $|\delta_2|/a^2$ is approximately an order of magnitude larger than $|\delta_1|/a$.

	\section{Comparison with known results and mass factorization}
		\label{sec:factorization}
		In the previous section, we calculated the differential cross section and studied the region of applicability of its expansion in the electron mass. We now compare our amplitude and cross section with known results and discuss the consequences of our calculation for the QED mass-factorization formula.
		
		The leading-power mass-factorization formula is derived from a factorized description of the amplitude in terms of hard, collinear, and soft momentum regions \cite{becher2007two}. Additional momentum regions can nevertheless arise in individual diagrams or master integrals \cite{smirnov1999problems,becher2007two,ter2024region}. These extra regions may cancel when the diagrams are combined into a gauge-invariant amplitude. To check whether they cancel in our case, we would need a complete region analysis. In our computational setup, this would involve an IBP-reduction of the amplitude in a parametric representation. Since such a reduction is not available here, we instead examine the structure of the final result and of the intermediate calculations.
		
		In the calculation of the massive master integrals in Ref.~\cite{lee2024master}, the expansion coefficients in the sequential limit  $m^2 \rightarrow 0$, followed by $T\rightarrow0$, were determined for certain integrals. As in the works~\cite{smirnov1999problems,ter2024region}, additional integration regions beyond the soft, hard, and collinear ones appear at the level of individual diagrams or integrals. In addition, the generalized power-series expansions of the master integrals \eqref{gps1} contain terms proportional to $(m^2)^{\lambda_i}$, with
		\begin{equation}
			\{\lambda_{1},\,\lambda_{2},\,\dots,\,\lambda_{7} \}  = \{0,-\frac{1}{2}  -  3\epsilon,-\frac{1}{2}-2\epsilon,-4\epsilon,-3\epsilon,-2\epsilon,-\epsilon\}. \label{lambdaagain}
		\end{equation}
	
		For a Feynman integral $j$, we use $[[j]_{m^a}]_{T^b}$ to denote the coefficient of $m^aT^b$ in its sequential expansion, first for $m^2\to0$ and then for $T\to0$. Although the expansion is expressed in powers of $m^2$, the half-integer parts of $\lambda_2$ and $\lambda_3$ generate odd powers of $m$. The coefficients associated with $\lambda_2$ are $[[j_{66} ]_{m^{2\lambda_2-4}}]_{T^{-3/2-\epsilon}}$ and $[[j_{74} ]_{m^{2\lambda_2-4}}]_{T^{1/2+\epsilon}}$, while those associated with  $\lambda_3$ are $[[j_{64} ]_{m^{2\lambda_3}}]_{T^{-1/2}}$ and $[[j_{73} ]_{m^{2\lambda_3}}]_{T^{-1/2}}$, where
		\begin{align}
			&j_{64} = \includegraphics[width=0.2\textwidth,align=c]{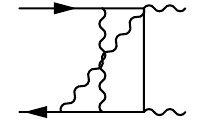}, \qquad 
			j_{73} = \includegraphics[width=0.2\textwidth,align=c]{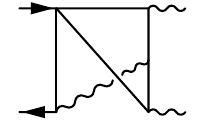}, \nonumber\\*  
			&j_{66} = \includegraphics[width=0.2\textwidth,align=c]{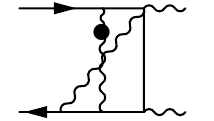}, \qquad 
			j_{74}  = \includegraphics[width=0.2\textwidth,align=c]{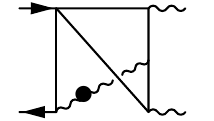}. \label{4diags1}
		\end{align}
		The dots on the diagrams in Eq.~\eqref{4diags1} denote squared propagators.
		
		At leading power in $m$, only terms with $\lambda_1=0$, $\lambda_6=-2\epsilon$, and $\lambda_7=-\epsilon$ survive in the two-loop gauge-invariant amplitudes \eqref{hviaa} and in the corresponding contribution to the differential cross section \eqref{ds00}. Terms associated with the remaining exponents cancel.
		
		The first subleading term in the two-loop amplitude is linear in $m$ and arises only from $\lambda_3$. The corresponding constants are purely imaginary, which makes the entire subleading term purely imaginary in our conventions. Since the Born amplitude is real, this term does not contribute to the interference between the Born and two-loop amplitudes entering the unpolarized differential cross section. Consequently, the first nonvanishing power correction to the two-loop contribution to the cross section is quadratic in $m$. At this order, as at leading power, only terms with $\lambda_1=0$, $\lambda_6=-2\epsilon$, and $\lambda_7=-\epsilon$ remain.
		
		A similar cancellation was found for the massive QED fermion form factor in Ref.~\cite{ter2024region}. There, additional regions arising in individual two-loop diagrams cancel in the complete form factor at both leading and next-to-leading power. Our result shows an analogous cancellation in the two-loop differential cross section for $e^+e^-\to\gamma\gamma$. 
		
		This simple pattern does not continue at higher orders. Starting at next-to-next-to-leading power, contributions associated with all exponents $\lambda_i$ are present. The cross section then also contains odd powers of $m$, beginning with $m^3$. Apart from the purely imaginary $\mathcal{O}(m)$ term discussed above, all contributions containing odd powers of $m$ in the two-loop amplitude and the differential cross section originate solely from $\lambda_2$. Their coefficients involve only rational functions of $T$ and square roots of $T$ and $1-T$.
		
		Work~\cite{fadin2023two} provides two-loop invariant amplitudes at leading power in the electron mass for $e^{+} e^{-} \rightarrow \gamma \gamma^{*}$, where one photon has a small virtuality. To compare these amplitudes with our results \eqref{hviaa}, we project both our tensor structures and those of Ref.~\cite{fadin2023two} onto a common auxiliary set of tensor structures. This allows us to relate the two bases and compare the amplitudes. Since the amplitudes in Ref.~\cite{fadin2023two} are expressed in terms of generalized polylogarithms with $T$-dependent indices, while our results are given in terms of classical polylogarithms, we perform the comparison numerically at many randomly selected kinematic points, repeating the evaluation at several working precisions. At a working precision of $p$ decimal digits, the modulus of the difference between the two amplitudes was below $10^{-p}$ at every point tested. At the maximum working precision of $1000$  digits, the difference was below $10^{-1000}$. This provides a stringent numerical check of leading-power QED mass factorization for this process.
		
		The one-loop result was obtained as a by-product of the main calculation and was also expressed in terms of massive master integrals. The resulting correction $\delta_1$ has the form of a generalized power series in the electron mass. Each term of this expansion agrees analytically with the corresponding term in the small-mass expansion of the exact result of Ref.~\cite{berends1973hard}. This analytic, term-by-term agreement provides an independent check of our one-loop result.

		A quantitative comparison with Ref.~\cite{engel2026low} would require numerical data from that calculation or an independent evaluation with \textsc{McMule} \cite{mcmule}. Since Ref.~\cite{engel2026low} presents the angular distribution only graphically and at a single energy, we restrict ourselves to a qualitative comparison. The published distribution is qualitatively consistent with our result.
	
		Two-loop QED corrections to Bhabha and M{\o}ller scattering have previously been obtained by direct calculations \cite{delto2023two}, but without a comparison with the mass-factorization formula. As far as we know, our work presents the first direct two-loop test of the QED mass factorization formula for a $2\to2$ process. Further investigation is needed to establish the general validity of the formula.
	\section{Ancillary files}
	\label{sec:ancillary}
		The analytic results obtained in this work are provided as ancillary material. For this purpose, we write the hard amplitudes \eqref{hviaa} as series in $m = \frac{m_e}{\sqrt{s}}$:
		\begin{equation}
			\mathrm{H}_k = \sum_{n=0}^{12} \mathrm{H}_{k,n+1}\,m^n,
			\qquad k=1,\dots,8, \label{file1}
		\end{equation}
		The supplementary material contains the following files:
		\begin{itemize}
			\item \texttt{Hkn/Hkn.l}, with $\texttt{l}=0,1,2$ corresponding to the Born, one-loop, and two-loop hard amplitudes \eqref{hviaa}, respectively. The amplitudes are stored in the form \eqref{file1} as nested lists indexed by $k$ and $n$. Only the two-loop file ($\texttt{l}=2$) contains classical polylogarithms, depending on $T=m^2-t/s$. The flags \texttt{nf} and \texttt{Lbl} identify the vacuum-polarization and light-by-light contributions shown in Figs.~\ref{fig:po} and \ref{fig:lbl}, respectively.
			\item \texttt{Glk}: the matrix \eqref{matr} constructed from the tensor structures.
			\item \texttt{W2V}: the combination $\mathcal{W}(\omega_0)+2\Re\mathcal{V}$ appearing in Eq.~\eqref{w2v}.
			\item \texttt{deltas/Born}, \texttt{deltas/delta1} and \texttt{deltas/delta2}: the Born, one-loop \eqref{d1form}, and two-loop \eqref{d2form} contributions to the unpolarized differential cross section, respectively, all stored as lists of the coefficients of $m^n$.
			\item \texttt{gammastar/Hn/Hnsmallqq.l}, with $\texttt{l}=0,1,2$: the hard amplitudes from Ref.~\cite{fadin2023two}, supplied as input for the numerical comparison with our results.
			\item \texttt{CSEvaluation.nb}: a \textsl{Mathematica} notebook for evaluating the hard amplitudes, constructing the unpolarized differential cross section, reproducing the plots presented in this paper, and performing the numerical comparison with Ref.~\cite{fadin2023two}.
		\end{itemize}
		The same ancillary material is available in the GitHub repository \url{https://github.com/VStotsky/eegg_highenergy/}.
	\section{Conclusion}
		In this paper, we have calculated the two-loop amplitude and the differential cross section for $l^+l^-\to\gamma\gamma$ in the high-energy, large-angle region $s,|t|,|u|\gg m_l^2$. The results are also applicable to the inverse process $\gamma\gamma\to l^+l^-$. The calculation includes corrections beyond the leading power in the lepton mass, thereby retaining mass effects that can be significant for accurate predictions. The region of reliable applicability is also examined numerically. For the expansion orders and accuracy considered in this work, our numerical analysis suggest that the results can be used for $s,|t|,|u|\gtrsim40m_l^2$.
		
		The decomposition into invariant amplitudes allows the results to be used for arbitrary polarizations of the external leptons and photons. We have expressed the amplitudes in terms of classical polylogarithms, making them suitable for efficient numerical evaluation.
		
		At leading power in the lepton mass, our two-loop amplitudes agree numerically with the results obtained from QED mass factorization. As far as we know, this is the first direct two-loop comparison with the mass factorization formula for a QED $2\to2$ process. 
		
		In future work, we plan to calculate the amplitude and differential cross section as an expansion near threshold, retaining successive power corrections in that expansion as well. Combining the resulting threshold expansion with the high-energy expansion obtained here will provide complementary descriptions of the two limiting regimes and the input needed to construct a theoretical prediction covering the full energy range.
	\section*{Acknowledgements}
		The author is grateful to Roman N. Lee for numerous discussions and guidance throughout this work, and to Petr A. Krachkov for his interest in the project.
		
	\appendix
	\section{Two-loop on-shell renormalization constants}
	\label{app:renorm}
	For completeness, we give here the explicit $\epsilon$-series for the two-loop coefficients $Z_{m,2}^{\mathrm{OS}}$ and $Z_{2,2}^{\mathrm{OS}}$ used in the construction of the counterterms:
	\begin{multline}
		Z_{m,2}^{\mathrm{OS}} = \tfrac{2 n_f+\tfrac{9}{2}}{\e ^2}
		+\tfrac{7 n_f+\tfrac{45}{4}}{\e }
		+\left(\tfrac{69}{2}-16 \zeta _2\right) n_f-30 \zeta _2-12 \zeta _3+48 \zeta _2 \ln{2}+\tfrac{199}{8}
		\\*
		+  \Big[ \left(-80 \zeta _2-56 \zeta _3+96 \zeta _2 \ln{2}+\tfrac{463}{4}\right)n_f -192 \operatorname{Li}_4(1/2) -165 \zeta _2-132 \zeta _3+252 \zeta _4
		\\*
		-96 \zeta _2 \ln^2{2}+288 \zeta _2 \ln{2}+\tfrac{677}{16}-8 \ln^4{2}\Big]\epsilon
		+\mathcal{O}\left(\epsilon^2\right),
	\end{multline}
	\begin{multline}
		Z_{2,2}^{\mathrm{OS}} = 
		\tfrac{4 n_f+\tfrac{9}{2}}{\e ^2}
		+\tfrac{\tfrac{19 n_f}{3}+\tfrac{51}{4}}{\e }
		+\left(\tfrac{1139}{18}-32 \zeta _2\right) n_f-78 \zeta _2-24 \zeta _3+96 \zeta _2 \ln {2}+\tfrac{433}{8}
		\\*
		+\Big[-384 \operatorname{Li}_4(1/2) + n_f \left(-152 \zeta _2-112 \zeta _3+192 \zeta _2 \ln {2}+\tfrac{20275}{108}\right)-267 \zeta _2-294 \zeta _3+504 \zeta _4
		\\*
		-192 \zeta _2 \ln ^2{2}
		+552 \zeta _2 \ln {2}+\tfrac{211}{16}-16 \ln ^4{2}\Big]\epsilon 
		+\mathcal{O}\left(\epsilon^2\right).
	\end{multline}
	
	\bibliographystyle{JHEP}
	\bibliography{eegg}
\end{document}